\documentclass[prb,twocolumn, superscriptaddress]{revtex4-1}
\usepackage{graphicx,amsmath,amssymb,amsxtra, color}
\renewcommand{\narrowtext}{\begin{multicols}{2} \global\columnwidth20.5pc}

\newcommand{\sign}{{\rm sign}}
\def\be{\begin{equation}}
\def\ee{\end{equation}}

\newcommand{\Eq}[1]{Eq.~(\ref{#1})}

\newcommand{\pab}{{\ensuremath{p_{\alpha\beta}}~}}
\begin{document}

\title{Gapless excitations in the Haldane-Rezayi state: The thin-torus limit}

\author{Alexander Seidel}
\affiliation{Department of Physics and Center for Materials Innovation, Washington University, St. Louis, MO 63136, USA}
\author{Kun Yang}
\affiliation{National High Magnetic Field Laboratory, Florida State
University, Tallahassee, Fl 32306, USA}

\date{\today}

\begin{abstract}
We study the thin-torus limit of the Haldane-Rezayi state.
Eight of the ten ground states are found to assume a simple product
form in this limit, as is known to be the case for many other quantum Hall
trial wave functions. The two remaining states have a somewhat unusual thin-torus limit,
where a ``broken'' pair of defects forming a singlet is completely delocalized.
We derive these limits from the wave functions on the cylinder, and deduce
the dominant matrix elements of the thin-torus hollow-core Hamiltonian. We find that
there are gapless excitations in the thin-torus limit.
This is in agreement with the expectation that local Hamiltonians stabilizing wave functions
associated with non-unitary conformal field theories are gapless. 
We also use the thin-torus analysis to obtain explicit counting formulas for the 
zero modes of the hollow-core Hamiltonian on the torus, as well as for the parent Hamiltonians
of several other paired and related quantum Hall states.
\end{abstract}

\maketitle

{\section{Introduction}

The theoretical study of electronic phases in the fractional 
quantum Hall regime owes much of its success to the
construction of analytic trial wave functions,\cite{laughlin1}
and their subsequent interpretation in a conformal field theory (CFT)
context.\cite{MR,naywil}
The use of CFT gives rise to powerful predictions regarding, e.g.,
the edge physics of a state or its statistics.
The connection between trial wave functions and CFT
allows for elegant derivation of results without further microscopic
studies of the wave functions themselves or their parent Hamiltonians.
The use of CFT in this way is, however, not free of conjecture. 
One exciting development in the field is the gradual improvement
of the foundation underlying these conjectures, e.g., regarding statistics.\cite{gurnay,read_viscosity, read_preprint2,bonderson}

Another important prediction based on the 
CFT correspondence holds that wave functions related to
non-unitary CFTs cannot be stabilized through local Hamiltonians
with an energy gap in their bulk excitation spectrum.\cite{gaffnian, read_chiral, read_viscosity,read_preprint2}
In this paper we will show that this statement is consistent
with a different scheme of attack recently found in the literature.
This is to take the thin-torus limit of a quantum Hall state,
and assume adiabatic continuity between this quasi one-dimensional (1D)
limit, and the limit of a two-dimensional (2D) torus without a ``thin'' dimension.
Thus far, this method has been mainly used to study systems with
a bulk energy gap.\cite{seidel1,seidel2,karlhede2,karlhede3}
Since early on, however, the assumption of adiabatic continuity
has also been made for systems with gapless bulk excitations.\cite{karlhede1}
More recently we have applied\cite{seidelyang} this method to a
possible transition\cite{readgreen} between the Halperin (331)-state\cite{halperin1} and the Moore-Read state\cite{MR}
triggered via interlayer tunneling. Here we apply the same approach to the 
Haldane-Rezayi (HR) state,\cite{HR} whose associated CFT is non-unitary.\cite{MR, wenwu, milo, gurarie97}

The thin-torus approach is itself based on a conjecture, 
namely that of adiabatic continuity
described above. The latter has led to some success in the past, 
but is so far less  well established overall compared to the CFT based conjecture.
Since the assumptions underlying the thin-torus method are, however, not field-theoretic in nature,
they may be viewed as quite independent of those used in the CFT approach.
It is thus reassuring that so far, the results of both approaches have been in
overall  agreement. This seems to include the issue
of quasi-particles statistics.\cite{seidel3, seidel_pfaff,flavin} 
The purpose of this paper is to demonstrate this agreement 
with regard to the existence of gapless excitations in the HR state.
Furthermore, we use the insights gained from the thin-torus limit to generalize
various counting formulas,\cite{read2, green} which have been given
for the zero energy modes of parent Hamiltonians
in the case of  spherical topology, to the torus.
This is possible since, in the thin-torus limit,
simple patterns generally appear that
lead to combinatorial principles which organize
the subspace of zero modes (cf. also Ref. \onlinecite{read1}).
These patterns may also be viewed as dominance
partitions of Jack polynomials,\cite{haldanebernevig2, haldanebernevig3}
and are related to ``patterns of zeros''.\cite{wenwang,wenwang2,barkeshli}

The paper is organized as follows.
In Section \ref{TT}, we take the formal thin-cylinder limit
of HR wave functions on the cylinder. This allows us to
infer all the patterns associated with topological sectors
in the thin-torus limit.
In Section \ref{modes} we discuss the matrix elements
of the thin-torus hollow-core Hamiltonian, and use this
to demonstrate the existence of gapless excitations
in the thin-torus limit.
In Section \ref{count} we use  
the combinatorial principles imposed
by the thin-torus Hamiltonian to derive torus
zero mode counting formulas 
for the HR state and various other paired and related
quantum Hall states.
We discuss our results in Section \ref{dis}.
Various technicalities relating to the 
thin-torus limit are presented in three Appendices. 

\section{The thin-torus limit\label{TT}}
\subsection{Preliminary considerations}

We focus on the Haldane-Rezayi state at filling factor $\nu=1/2$.
This is the top state in the HR sequence, and is its most elementary state
in terms of torus degeneracy. Even so, for even particle number,
the torus degeneracy of 10 of this state is already rather large.
The corresponding torus wave functions have been constructed
 in Refs. \onlinecite{KeskiVakkuri, read2}.
The full set of patterns associated with these states 
have been known for some time, \cite{KITP_talk, Stockholm_talk}
 and naturally appear 
when the thin-torus limit is taken.
In this limit, trial wave functions for ground states usually approach
simple product states,\cite{rezayi94} and patterns appear as patterns
of occupancy numbers of consecutive lowest-Landau-level (LLL)
orbitals. For the ten HR ground states,
 eight of the associated patterns are easy to guess. 
The particles 
in the associated product state
carry a spin or pseudo-spin label, and must form a singlet, 
since the HR state is a two-component singlet state.
A simple pattern, that is, one with a simple
unit cell, must therefore contain at least two particles per unit cell that form a singlet.
Natural candidates are thus given by the following patterns:
\begin{subequations}\label{AB}
\begin{align}
\includegraphics[width=5.5cm]{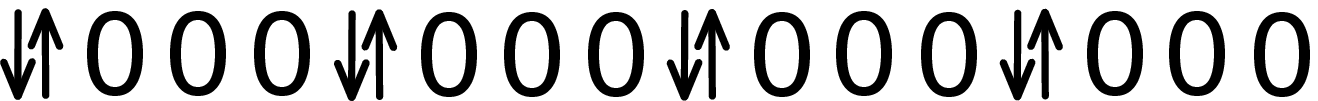}\label{A}\\
\includegraphics[width=5.5cm]{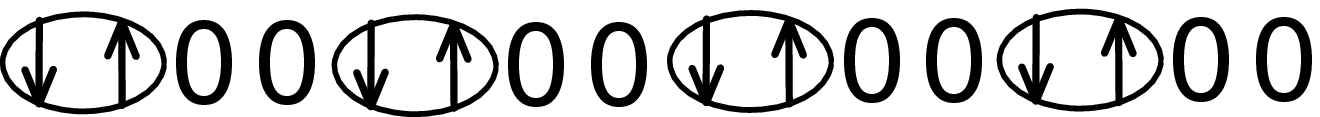}\label{B}
\end{align}
\end{subequations}
which we call patterns $A$ and $B$.
Here, as usual, patterns denote consecutive occupancies of LLL orbitals in the thin-torus limit
and ovals denote singlets formed by two electrons in different orbitals. 
Taking into account
translational symmetry on the torus, these patterns together account for eight degenerate
ground states. They have also appeared in the literature more recently from a 
``generalized Pauli principle'' point of view\cite{ronny, bernevig} (cf. also Ref. \onlinecite{milo09}). 
The above, however, leaves the thin-torus limits of two more ground states undetermined,
and the associated patterns have to our knowledge not yet appeared in the printed
literature.\cite{KITP_talk, Stockholm_talk} 
Interestingly, these patterns are not ``simple,'' in the sense that they are not described by a 
small unit cell. Indeed, since any such unit cell must harbor at least two particles in a singlet state,
and hence must at least consist of four LLL orbitals at $\nu=1/2$, any such pattern would imply the existence of four more ground-states. It is therefore not possible that the thin-torus limit of the
remaining two ground states yields a pattern with a simple unit cell.
In fact, as we will show below, it is the existence of these two rather ``special'' ground-state sectors
which  implies the existence of gapless excitations in the state, at least in the thin-torus limit.

{\subsection{The $A$ pattern}

We proceed by first deriving the thin-torus limits of \Eq{AB} in a systematic way,
by following the general method developed earlier by us.\cite{seidelyang}
In this method, the thin-torus pattern is first obtained on a ``squeezed lattice''
and then unsqueezed by well-defined rules. Here we give a
self-contained account of this method, including some details not explicit in the original work.

Since torus wave functions are generally complicated and lack a simple polynomial structure, we
use  the fact that all torus patterns can also be obtained by working on the cylinder with appropriate
``boundary conditions,'' corresponding to various edge state configurations (vacua) (cf. Ref. \onlinecite{seidelyang}). 
We will demonstrate this as we go along.
To begin, we take the polynomial structure associated with the HR ground state in planar or
spherical geometry\cite{HR} and write down a corresponding LLL state on the cylinder:
\be\label{HRA}
\begin{split}
H_A(\{\xi_\alpha\})= \sum_{\sigma\in S_{N/2}}\cfrac{(-1)^\sigma}{(\xi_{1\downarrow}-\xi_{\sigma_1\uparrow})^2\dotsc(\xi_{\frac{N}{2}\downarrow}-\xi_{\sigma_\frac{N}{2}\uparrow})^2}\\
\times\,\prod_{\alpha<\beta}(\xi_\alpha-\xi_\beta)^2\,.
\end{split}
\ee
Here, $\xi_{\alpha}=\exp(\kappa(x_\alpha+iy_\alpha))$, $\kappa=2\pi/L_y$ is the inverse radius
of the cylinder, greek letters are multi-indices of the form 
$\alpha=(i,S_z)$, referring to the $i-th$ particle with spin projection $S_z=\uparrow$
or $S_z=\downarrow$, and $\alpha<\beta$ refers, for definiteness, to the lexicographic order 
$(1,\downarrow)<\dotsc<(N/2,\downarrow)<(1,\uparrow)<\dotsc<(N/2,\uparrow)$.
 There are $N/2$ particles for each projection.
The structure of \Eq{HRA}
has the familiar form of a Laughlin-Jastrow factor multiplied by a determinant,
which we have written out explicitly. We suppress the 
standard Gaussian factor $\exp(-\frac{1}{2}\sum_\alpha x_\alpha^2)$ here and in the
following.

One can now imagine expanding \Eq{HRA} into polynomials of the form
$\prod_\alpha \xi_\alpha^{n_\alpha}$. Let $C_{\{n_\alpha\}}$ be the 
coefficient of such a monomial.
This coefficient is essentially the amplitude $A_{\{n_\alpha\}}$ for the 
product state that has particles occupying the LLL 
orbitals $\{n_\alpha\}$, except for a normalization factor:\cite{rezayi94}
\be\label{amplitude}
  A_{\{n_\alpha\}}=e^{\frac{1}{2}\kappa^2 S} \,C_{\{n_\alpha\}}\;,
\ee
where
\be\label{S}
S=\sum_\alpha n_\alpha^2\;.
\ee
For this reason, those product states 
will dominate the thin-cylinder limit $\kappa\rightarrow\infty$
for which the quantity $S$ takes on the maximum value, subject
to the constraint that the associated polynomial coefficients
$C_{\{n_\alpha\}}$  are non-zero.
The $n_\alpha$ in any monomial with non-zero coefficient
are of the following form: 
\be\label{na}
n_\alpha=\sum_{\beta\neq\alpha} (m_{\alpha\beta}+p_{\alpha\beta})\,,
\ee
where $m_{\alpha\beta} $ and $p_{\alpha\beta}$ 
are symmetric and anti-symmetric matrices, respectively,
that correspond to the selection of the term $\xi_\alpha^{m_{\alpha\beta+\pab}}
\xi_\beta^{m_{\alpha\beta-\pab}}$ in a factor depending on
$\xi_\alpha$ and $\xi_\beta$ in \Eq{HRA}.
In the present case, all  factors surviving cancellation
are of the form $(\xi_\alpha-\xi_\beta)^2$.
For these terms, $m_{\alpha\beta}=1$ and
$p_{\alpha\beta}\in\{-1,0,1\}$.
For a fixed permutation $\sigma$, however,
factors corresponding to certain pairs $(\alpha\beta)$ in the Laughlin-Jastrow factor
are canceled, and this imposes additional constraints on  
$m_{\alpha\beta} $ and $p_{\alpha\beta}$.
More precisely, 
\be\label{sconstraint}
\begin{split}
&m_{\alpha\beta}=0 \mbox{ for } \alpha=\beta\\ 
 &m_{\alpha\beta}=0 \;\mbox{for}\, (\alpha,\beta)\in \{ (i\downarrow, \sigma_i\uparrow): i=1\dotsc N/2\}\\
 & m_{\alpha\beta}=1\;\mbox{otherwise.}
\end{split}
\ee
where $(\cdot,\cdot)$ denotes an unordered pair, and
\be\label{prange}
\pab\in\{-m_{\alpha\beta},-m_{\alpha\beta}+1\dotsc m_{\alpha\beta}\}\;.
\ee
A given permutation $\sigma$ thus induces a
pairing between up-spins and down-spins, which determines
the possible monomials $\prod_\alpha \xi_\alpha^{n_\alpha}$
through the rules \eqref{na}-\eqref{prange}.
In addition, $p_{\alpha\beta}$ is constrained by its
anti-symmetry.
Making all possible choices for $p_{\alpha\beta}$
generates all possible monomials for a
given pairing $\sigma$.
In Appendix \ref{app1}, we show that
in order to maximize \Eq{S}, \pab must be of the following form:
\be\label{pab}
\pab= m_{\alpha\beta}\,\sign(\rho_\alpha-\rho_\beta),
\ee
where $\rho\in S_N$ is a permutation of $N$ objects.
This means that $|\pab|$ must always be chosen to have its
maximum possible value, $m_{\alpha\beta}$, and the sign
structure is determined by an ordering of the particle indices.
\Eq{pab} implies that in each factor $(\xi_\alpha-\xi_\beta)^2$, we pick either
the term $\xi_\alpha^2$ or the term $\xi_\beta^2$, as determined by $\rho$,
but never the mixed term.
$\rho$ may be thought of as determining the order of
the $N$ particles
on a ``squeezed'' lattice of $N$ sites. The term ``squeezed''
implies that the arrangement of the particles on the squeezed
lattices is a precursor of the thin-cylinder pattern
we seek, but with all inter-particle distances set equal to 1
(hence, with vacant sites ``squeezed out'').
The corresponding un-squeezed pattern, the thin-cylinder limit
of the state, is then obtained by evaluating \Eq{na} for given
\pab  and $m_{\alpha\beta}$, as
determined by $\sigma$, $\rho$ via Eqs. \eqref{sconstraint} and \eqref{pab}. 
The resulting $n_\alpha$ determine the exponents in a dominant
monomial $\prod_\alpha \xi_\alpha^{n_\alpha}$, which corresponds
to a product state where the particle with index $\alpha$ occupies the
LLL orbital with index $n_\alpha$.

In the above, the choice of the pairing $\sigma$ is arbitrary, since
varying $\sigma$ only anti-symmetrizes the resulting monomials.
We may thus choose $\sigma=id$ without loss of generality.
However, for given $\sigma$, the proper choice of $\rho$ that maximizes
\Eq{S} is not arbitrary. Rather, to determine the proper arrangement of particles
on the squeezed lattice, we must minimize an effective energy, as done
in Appendix \ref{app2}. This yields that, for $\sigma=id$,
a proper choice of $\rho$ is
\be\label{rho}
\rho= 1\downarrow, 1\uparrow,  2\downarrow, 2\uparrow,\dotsc, \frac{N}{2}\downarrow, \frac{N}{2}\uparrow\;.
\ee
That is, the particles paired according to $\sigma$ must be neighbors on the
squeezed lattice. 
{Permuting neighboring up- and
down-spin particles with the same particle index results in the same
$p$-matrix, since $\pab=0$ for the corresponding multi-indices $\alpha$, $\beta$.}
We may of course permute the overall order of pairs, where different orders
lead to different monomials contributing to the single Slater determinant
that dominates the thin-cylinder limit.
Using \Eq{rho} in \Eq{na}, 
the resulting values $n_\alpha$ then are, listed in the order given by the right-hand side of \Eq{rho}:
\be
n_{\alpha}= 0,0,4,4,8,8,\dotsc\,.
\ee
When the corresponding spins are distributed over LLL orbitals accordingly, this 
yields the pattern \Eq{A}. 
It follows from the above that each monomial that is present in the
associated Slater determinant is obtained from one and only one 
choice for the $p$-matrix.\footnote{Since all these monomials maximize
$S$, \pab must be of the form \eqref{pab} with $\rho$ 
arranging the pairs of $\sigma$ into neighbors on the squeezed lattice.
As explained in the text, choices for $\rho$ that lead to different \pab
also lead to different monomials, related by index permutation.} 
The corresponding monomial is hence 
generated in only one way, and is thus guaranteed to have a non-zero coefficient.

\subsection{The $B$ patterns}

\begin{figure}
\includegraphics[width=7cm]{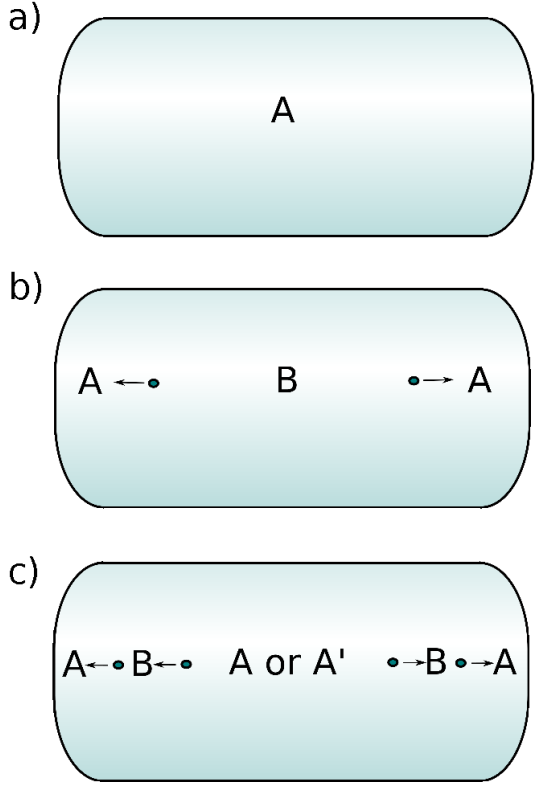}
\caption{Transitions 
between different ground state sectors of the HR state by quasi-hole insertion.
We start with the polynomial \Eq{HRA}, which corresponds to the unique ground state on the sphere, and consider the associated state on the cylinder (a).
The thin-cylinder limit of this state is given by the $A$ pattern, \Eq{A}.
Insertion of two quasi-holes (b) leads to a state whose thin-cylinder limit
is shown in Fig. \ref{domain}a), with a $B$ pattern appearing inside the 
$A$ pattern, and separated from it by two domain walls.
When the quasi-hole positions are taken to $\pm\infty$, the state
approaches \Eq{HRB}, whose thin-cylinder limit
is described by the $B$ pattern \eqref{B}.
We may iterate the process by inserting a second quasi-hole pair (c).
For this one has the choice of breaking a pair in the
pairing wave function of the state, or not.\cite{read2}
In the latter case, another $A$-pattern will appear in the
thin-cylinder limit, surrounded by $B$-patterns. If, on the other hand, the pair is broken,
one obtains an ``$A'$-string'' instead. With the quasi-holes taken to 
$\pm\infty$, the state approaches the wave function \Eq{HRC}. Its 
thin-cylinder limit is the superposition of states 
represented in Fig. \ref{Aprime},
which features a delocalized broken pair, and
which
we will loosely
refer to as an $A'$-string. (See also Fig. \ref{domain}(e) for the appearance of an $A'$-string
terminating in domain walls before quasi-holes are taken to $\pm\infty$.) 
\label{holecreation}}
\end{figure}
The $N$-particle ground state on an infinite cylinder is infinitely degenerate by
translational symmetry. However, valid incompressible ground state wave functions
come in a finite number of different classes, corresponding to the different patterns obtained
when taking the thin-cylinder limit.  To obtain a different class of ground state,
we may use the following procedure. The state \eqref{HRA} describes a
``ribbon'' of incompressible fluid with two opposing edges.\cite{rezayi94} Inside this ribbon,
we can make two quasihole-type excitations (Fig. \ref{holecreation}). One member of the pair 
is then formally taken  across the left edge to $x=-\infty$, whereas the other is taken
across the right edge to $x=+\infty$. With the holes removed, the state has healed into
an incompressible fluid, but in a different ground state sector not (necessarily) related to the original one by translation. On the sphere, the same process corresponds to placing the two members
of the quasi-hole pair one at the north and one at the south pole. We may thus 
take the polynomials associated with quasi-hole wave functions from work
done in spherical geometry.\cite{read2} Recall that the polynomial
\eqref{HRA} corresponds to the unique ground state on the sphere.
As observed in Ref. \onlinecite{read2}, there is only one class
of two-quasi-hole states that can be generated inside this ground state,
whereas for $2n$ holes with $n>1$, various classes (sectors) can be 
distinguished due to additional degrees of freedom corresponding to
pair breaking. We will find a natural interpretation for this behavior in the
quasi 1D limit.
We take  the unique class of polynomial wave functions describing
two quasi-holes generated in the ground state \eqref{HRA}, parametrized
by the two quasi-hole coordinates $h_1$ and $h_2$, and take the
limits described above. The resulting ground state polynomial is then
\be\label{HRB}
\begin{split}
H_B(\{\xi_\alpha\})= \sum_{\sigma\in S_{N/2}}\cfrac{(-1)^\sigma
(\xi_{1\downarrow}+\xi_{\sigma_1\uparrow})\dotsc(\xi_{\frac{N}{2}\downarrow}+\xi_{\sigma_\frac{N}{2}\uparrow})
}{(\xi_{1\downarrow}-\xi_{\sigma_1\uparrow})^2\dotsc(\xi_{\frac{N}{2}\downarrow}-\xi_{\sigma_\frac{N}{2}\uparrow})^2}\\
\times\prod_{\alpha<\beta}(\xi_\alpha-\xi_\beta)^2\,.
\end{split}
\ee
It is not difficult to see that, like $H_A$, $H_B$ has the analytic properties that 
render it a zero-energy eigenstate of the hollow-core Hamiltonian.\cite{HR,read2}
The fact that $H_B$ describes an incompressible state at the same filling factor as
$H_A$ will be apparent in the thin-cylinder limit.

It remains to see how the rules of the game described above to find the 
dominant thin-cylinder monomials of $H_A$ change for the polynomial 
$H_B$. It is easy to see that only the central 
line of \Eq{sconstraint} needs
modification:
\be\label{sconstraint2}
\begin{split}
  m_{\alpha\beta}=1/2, \;\mbox{for}\, (\alpha,\beta)\in \{ (i\downarrow, \sigma_i\uparrow): i=1\dotsc N/2\}\;.
\end{split}
\ee
The change just describes the following fact: Now 
terms in the Laughlin-Jastrow factor
that belong to the pairing associated with $\sigma$ are not just canceled, but are replaced by
$(\xi_\alpha+\xi_\beta)$.
For the reasons described above, we can look at $\sigma=id$ without loss of generality.
\Eq{rho} is then still a solution for $\rho$ that maximizes $S$, \Eq{S} (Appendix \ref{app2}).
The resulting monomial corresponds to the product state
\be\label{Bpattern}
\includegraphics[width=6cm]{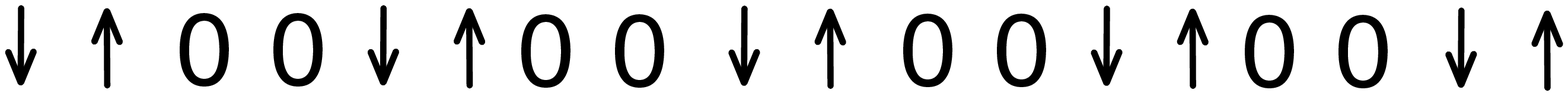}
\ee
However, permuting the order of a pair $i\uparrow, i\downarrow$ in $\rho$,
\Eq{rho}, now leads to a different \pab that also maximizes $S$, with the members
of the 
$i$-th pair in \Eq{Bpattern} trading places.
It follows that the dominant state in the thin-cylinder limit is an equal-amplitude
superposition of all states generated from \Eq{Bpattern} by exchanging 
the opposite spins of neighboring pairs in all possible ways.
It is worth checking that the signs work out for this state to describe
a product of singlets, as shown in \Eq{B}.
For this it is enough to consider a state with $N=2$ particles.
According to the above, the dominant term in the thin-cylinder limit
is
\be\label{singlet}
 \xi_\alpha^0 \xi_\beta^1 +  \xi_\alpha^1 \xi_\beta^0
= \xi_\alpha+\xi_\beta\;.
\ee
Here, $\alpha$ and $\beta$ refer to the down-spin
and the up-spin particle, respectively.
In writing down the wave functions \eqref{HRA}, \eqref{HRB},
we have distinguished two sub-species of opposite spin
and have not imposed any anti-symmetry condition between these distinguishable
sub-species. Rather, in \Eq{singlet} $\alpha$ is always associated with
a down-spin particle, and $\beta$ is always associated with an up-spin
particle. Viewed as a function of position {\em and} spin coordinates,
\Eq{singlet} should be multiplied by the spinor 
$\delta_{s_\alpha,\downarrow}\delta_{s_\beta,\uparrow}$.
The fully anti-symmetrized wave function is then
\be
( \xi_\alpha\delta_{s_\alpha,\downarrow}\delta_{s_\beta,\uparrow}
-\xi_\beta\delta_{s_\alpha,\uparrow}\delta_{s_\beta,\downarrow})
-( \xi_\alpha\delta_{s_\alpha,\uparrow}\delta_{s_\beta,\downarrow}
-\xi_\beta\delta_{s_\alpha,\downarrow}\delta_{s_\beta,\uparrow})\,.
\ee
This is easily seen to be the difference between two Slater
determinants, describing a singlet. Hence, despite the
seeming lack of anti-symmetry, \Eq{singlet} describes a singlet, 
and we recover the pattern \Eq{B} in the thin-cylinder limit of 
\Eq{HRB}. 
The fact that the common coefficient of the dominant monomials obtained here
is  non-zero follows from observations similar to those made at the end of the preceding section.

\subsection{The $A'$ pattern\label{Ap}}
\begin{figure}
\includegraphics[width=8cm]{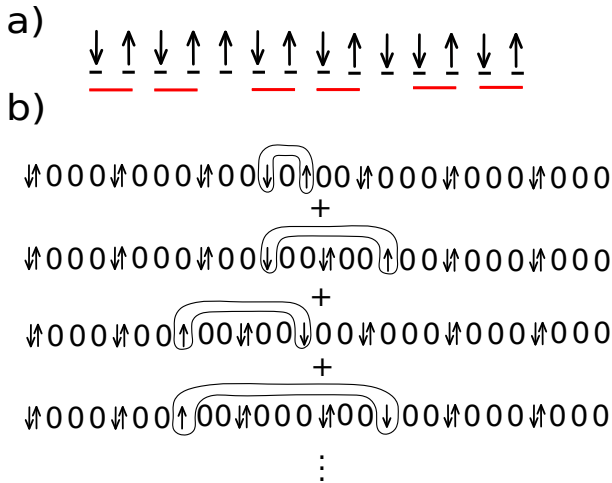}
\caption{
The thin-cylinder limit of the $A'$ ground state, \Eq{HRC}.
(a) One of many squeezed lattice configurations, corresponding
to a choice of the permutation $\rho$ that maximizes the quantity
$S$, \Eq{S}. A particular choice of $\rho$ corresponds to a squeezed lattice
configuration where each spin carries a particle number index. 
This is not shown, since permutation of like-spin indices does not
change the value of $S$, as must be the case by the anti-symmetry of the 
wave function. The short underscores represent squeezed lattice sites,
and the long underscores indicate the pairing $\cal P$, \Eq{mab2}.
Pairs must be nearest neighbors on the squeezed lattice, 
whereas the positions of the members of the broken pair
are arbitrary.
Together, $\rho$ and $\cal P$ completely define the $p$-matrix,
\Eq{mab}.
(b) The thin-cylinder limit of the $A'$-state,
obtained as an equal-amplitude superposition of all product states
derived by unsqueezing all squeezed lattice configurations
with given pairings of the form described under (a). 
This is done by using the associated $p$-matrix described in (a)
to form monomials according to Eqs.
\eqref{mAp}, \eqref{na}.
The unsqueezed version of (a) is shown in the last line.
As a result, the state is one where a 
completely {\em delocalized}
pair of charge neutral defects
forming a singlet separates two mutually out-of-phase $A$-patterns.
\label{Aprime}}
\end{figure}
The missing ground state sectors can be obtained by iterating the above procedure.
We thus create two more quasi-holes in the state \eqref{HRB}.
The number of inequivalent choices for this is the number of inequivalent
choices to make four quasi-holes in the original ground state \eqref{HRA},
which is in one-to-one correspondence with the ground state on the sphere.
For this it was found in Ref. \onlinecite{read2} that there are two such possibilities:
One may be described as having a ``broken pair'', while the other does not.
In the latter case, after sending both  quasi-hole pairs off to $\pm\infty$,
we obtain a state whose thin-cylinder limit is again of the $A$-type, \Eq{A}.
Hence we focus on the former case, where the state has a broken pair.
The analytic form of these states is more complicated, especially so on the torus.\cite{KeskiVakkuri,read2}
On the cylinder, again after sending the members of each pair to $\pm\infty$, 
we obtain the following wave function:
\be\label{HRC}
\begin{split}
H_C(\{\xi_\alpha\})= \sum_{\sigma\in S_{N/2}}
 \sum_{\lambda\in S_{N/2}}
(-1)^\sigma (-1)^\lambda
\prod_{i=2}^{N/2}\cfrac{(\xi_{\sigma_{i}\downarrow}+\xi_{\lambda_{i}\uparrow})^2}{(\xi_{\sigma_{i}\downarrow}-\xi_{\lambda_{i}\uparrow})^2}\\
\times\prod_{\alpha<\beta}(\xi_\alpha-\xi_\beta)^2\,.
\end{split}
\ee
In the above expression, the pairing between up-spin and down-spin particles is now
facilitated two permutations $\lambda$ and $\sigma$. The pairs thus formed are of the form
$(\sigma_i \downarrow, \lambda_i \uparrow)$, where the pair with index $i=1$ does not appear
in the pairing factor of the state, and can hence be thought of as ``broken''.
The thin-cylinder analysis is seemingly simple. In \Eq{HRC}, the summand corresponding to a
fixed pairing $\sigma$, $\lambda$ is of the form
\be\label{pseudoLJ}
\prod_{\alpha<\beta} (\xi_\alpha\pm \xi_\beta)^2\,,
\ee
where in each factor, the sign depends on $\alpha$, $\beta$, and the pairing.
The dominant monomials in such a term are the same as those
in a pure Laughlin-Jastrow factor. That is, 
the problem of finding the dominant monomial in the above product
again reduces to \Eq{na}, this time with 
\be\label{mAp}
m_{\alpha\beta}=1-\delta_{\alpha\beta}
\ee
(all factors in the product are homogeneous and of the same order).
The analysis given above again
implies
that the dominant monomials of \Eq{pseudoLJ} never make use of the
mixed term in $(\xi_\alpha\pm \xi_\beta)^2$. These monomials would
give rise to states where every second LLL-orbital is occupied, in all possible
ways. However, the monomials thus generated do not depend
on the permutations $\lambda$, $\sigma$ at all, since only the
mixed term in  $(\xi_\alpha\pm \xi_\beta)^2$ depends on the latter.
It is thus clear that in the sum over $\lambda$, $\sigma$ in \Eq{HRC}, 
all these naively dominant monomials cancel.

We thus have to look for those monomials in \Eq{pseudoLJ}
with the largest value of $S$, \Eq{S}, whose coefficient remains non-zero
after the sum over $\lambda$, $\sigma$ is taken in \Eq{HRC}.
Note that, since $\lambda$, $\sigma$ only affect the signs in \Eq{pseudoLJ},
the possible range of values for \pab is always $\{-1,0,1\}$, independent of $\alpha$, $\beta$.
Each choice for \pab corresponds to the choice of the term
proportional to $\xi_\alpha^{1+\pab} \xi_\beta^{1-\pab}$ in the factor
$(\xi_\alpha\pm \xi_\beta)^2$. It is thus useful to think of \Eq{HRC}
in terms of two {\em independent} sums:
\be
\begin{split}\label{coeff}
H_C(\{\xi_\alpha\})= \sum_{\{p_{\alpha\beta}\}}
\prod_{\alpha<\beta} \xi_\alpha^{1+\pab} \xi_\beta^{1-\pab}\\
\times  \sum_{\sigma,\lambda\in S_{N/2}}
 (-1)^\sigma (-1)^\lambda \mbox{coeff}(\lambda,\sigma,\{p_{\alpha\beta}\}) \\
\end{split}
\ee
The problem is thus to choose \pab such that \Eq{S} is maximized, subject
to the constraint that the second line in the above equation does not vanish.
\footnote{In general, different $p$-matrices could contribute to the same monomial.
This, however, will not be the case for the dominant ones, as shown below.}
Suppose, now, that \pab is such that for the $i$-th down-spin particle,
 we have
\be\label{unpaired}
 p_{i\downarrow,\beta}=- p_{\beta, i\downarrow}=\pm 1 \;\; \mbox{for all }\beta \; .
\ee
We will now show that there can be only one such $i$ in order for the sum over coefficients
in \Eq{coeff} not to vanish. If there were two indices $i_1$ and $i_2$ that satisfy \Eq{unpaired},
then exchanging the positions $\sigma^{-1}(i_{1,2})$ 
of these two indices in the permutation $\sigma$ 
changes the sign  $(-1)^\sigma$, but leave the coefficient $\mbox{coeff}(\sigma,\lambda,\{p_{\alpha\beta}\})$ unchanged. This is so because this coefficient can depend on 
$\sigma^{-1}(i)$,  only through the mixed term in 
$(\xi_{i\downarrow}\pm \xi_\beta)^2$ (whose sign depends on whether $\beta=(\lambda_{\sigma^{-1}(i)},\uparrow)$ or not), but for $i_1$, $i_2$ satisfying \eqref{unpaired}, these mixed terms do not enter
the monomial associated with \pab. Therefore, we can have at most one down-spin index
$i$ satisfying \eqref{unpaired}. For the same reasons there can be at most one up-spin index
$j$ with 
\be\label{unpaired2}
 p_{j\uparrow,\beta}=- p_{\beta, j\uparrow}=\pm 1 \;\; \mbox{for all }\beta \; .
\ee
When \pab is subject to the additional constraint that at most one  $i$ and one $j$ satisfy
\eqref{unpaired} and \eqref{unpaired2}, respectively, we can still use the arguments of Appendix
\ref{app1} to show that the \pab maximizing \Eq{S} is of the form
\be\label{mab}
\pab = s_{\alpha\beta}\,\sign(\rho_\alpha-\rho_\beta),
\ee
for some permutation $\rho$, only we cannot always choose the maximum value
$m_{\alpha\beta}= 1$ for $|p_{\alpha\beta}|=s_{\alpha\beta}$ any more.
We will, however, still want to choose $s_{\alpha\beta}=1$ as often as possible,
in order to approach the unconstrained maximum value of $S$ as closely as possible.
To do this, we choose one index $(i\downarrow)$ and one index  $(j\uparrow)$ for which \eqref{unpaired} and \eqref{unpaired2}
will be satisfied. The remaining $N-2$ particle indices 
are organized into a pairing ${\cal P}= \{(\alpha,\beta)\}$ ({\em different} from the pairing
induced by $\sigma$ and $\lambda$), each pair 
consisting of an up-spin and a down-spin index.
We then define
\be\begin{split}\label{mab2}
&s_{\alpha\beta}=0 \mbox{ if } \alpha=\beta\mbox{ or }(\alpha,\beta)\in {\cal P}\\
&s_{\alpha\beta}=1 \mbox{ otherwise}
\end{split}
\ee
It is clear that this assigns a minimum number of off-diagonal zeros to the
$s$-matrix, subject to the constraints described above. In any
monomial obtained from Eqs. \eqref{mab}, \eqref{mab2}, 
the indices $(i\downarrow)$ and $(j\uparrow)$ indeed play
the role of a ``broken pair'', since no term coming from a factor
$(\xi_{i\downarrow}\pm \xi_\beta)^2$ or $(\xi_{j\uparrow}\pm \xi_\beta)^2$
is ever affected by the relative sign that comes from the pairing wave function.

The permutation $\rho$ that leads to a maximum value of $S$ is not 
arbitrary, but its proper choice depends somewhat on the pairing $\cal P$. 
Again, we consider $\rho$ to define a squeezed lattice configuration as described above.
Then, as we show in Appendix \ref{app2}, we must arrange the pairs in $\cal P$ as
nearest neighbors on the squeezed lattice. The location of the members of the broken
pair is, however, arbitrary. This leads to the possible squeezed lattice configurations
shown in Fig. \ref{Aprime}(a). The corresponding unsqueezed configurations,
obtained by plugging the result into \Eq{na}, are shown in Fig. \ref{Aprime}(b)
(last line).
As shown in the figure, there are still many other 
different degenerate configurations,
corresponding to all possible positions of the members of the broken pairs.
The term dominating the thin-cylinder limit  of \Eq{HRC} is an equal-amplitude superposition
of all the corresponding monomials. In Appendix \ref{app3}, we evaluate the coefficient
of these monomials to be $2^{N-2} (N/2-1)!$, up to a sign, showing in particular that it is non-zero.
Moreover, the superposition is symmetric in the positions of the members of the broken pair.
By the same argument given at the end of the preceding section, 
this implies that this broken pair forms a singlet, as it should.
The members of the remaining, unbroken pairs each occupy the same orbital,
and thus form singlets automatically. 

As is clear from the graphical representation
in Fig. \ref{Aprime}(b), the thin-cylinder limit of \Eq{HRC}
is an equal-amplitude superposition of all states where a delocalized singlet 
separates two mutually out-of-phase $A$-patterns. More precisely, the $A$-patterns
thus separated differ by a shift of two orbitals.
On the torus, there are exactly two such states, related by a single translation.
Indeed, one will easily see that translating the state 
displayed in Fig. \ref{Aprime}(b) by {\em two} orbitals
leaves the state invariant, whereas a single translation
leads to the appearance of the other pair out of 
the four possible $A$-patterns, separated by defects.
We will henceforth refer to these thin-torus/cylinder states as $A'$-states or -patterns.
It is natural to think of these states not as defining new topological sectors,
but as $A$-type states in the presence of a single ``zero-energy excitation.''
We further elaborate on this notion in the following section.

\section{Properties of the thin-torus hollow-core Hamiltonian: Matrix elements and
gapless excitations\label{modes}}

This section presents the case for gapless excitations in the thin-torus limit.
It details the logic behind results first given
in Refs. \onlinecite{KITP_talk, Stockholm_talk}. The central step is the correct extraction of the dominant matrix
elements of the hollow-core Hamiltonian in the thin-torus limit. The diagonal parts of these matrix
elements give rise to what has more recently been proposed
as a ``generalized Pauli principle'' applying to the HR state.\cite{bernevig}
We find it crucial, however, to also pay attention to off-diagonal matrix elements
that survive in this limit. This  then implies existence of gapless modes
in the thin-torus limit. Our results will also allow us to derive a detailed formula for the number of
zero energy eigenstates, or
``zero modes'', of the hollow-core Hamiltonian on the torus in the presence of quasi-holes.

It is well known that the analytic properties of the HR-states render them unique
zero-energy ground states of the hollow-core Hamiltonian.\cite{HR}
At filling factor $\nu=1/2$, there are ten such ground states on the torus,\cite{read2, KeskiVakkuri}
irrespective of the aspect ratio of the torus.
In the thin-torus limit, these ten ground states evolve into product states
corresponding to the
patterns that we have seen to emerge in the thin-cylinder limit.
Indeed, when expressed as a second-quantized Hamiltonian
in the LLL-basis, the hollow-core Hamiltonian will
be dominated by highly local terms, with sub-dominant terms
exponentially suppressed (cf., e.g., Ref. \onlinecite{seidel1}). These dominant terms are identical in the case of a thin cylinder and a thin torus, for a given size (perimeter) of the
small dimension.
In general, 
the extraction of the proper dominant terms,
which directly imply the thin-torus limiting states,
requires the use of degenerate perturbation theory, as we described earlier.\cite{seidelyang}
Here we pursue a shortcut instead. The dominant terms of the thin-torus
Hamiltonian follow quite unequivocally from the knowledge of the limiting zero-energy ground states,
and the fact that the limiting Hamiltonian is local, with terms decaying exponentially
with distance.
This decay happens over a characteristic scale that is {\em short} compared to the orbital
separation. The latter essentially implies that no two 
terms can compete (leading to cancellations in the ground state energy) 
that act at a different range.
This imposes severe constraints on the effective thin-torus Hamiltonian.

Considering the $A$- and $B$- ground states, clearly there must be terms 
in the thin-torus Hamiltonian that
assign an energy to any four adjacent
orbitals containing more than two particles, or else there would be
zero modes at higher filling factor.
Likewise, there must be an energy cost associated with any two particles
that are no more than two orbitals apart, 
{\em if} these two particles
are forming a triplet.
 There is no such energy cost for singlets.
Without this triplet energy, we could make many $\nu=1/2$ zero modes
by placing one particle of any spin orientation in every second orbital.
In the presence of only these two types of interactions, 
the $A$- and $B$-type product states would be the unique
zero energy ground states at filling factor $\nu=1/2$.  
\begin{figure}
\includegraphics[width=8cm]{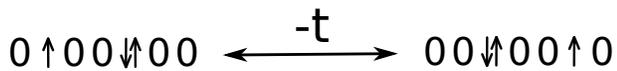}
\caption{
The off-diagonal matrix element present in the thin-torus hollow-core
Hamiltonian. This, together with the ``detailed balance'' condition
\Eq{Vt}, ensures that states containing an $A'$-type string 
with one or two delocalized defects are zero energy states in the
thin-torus limit.
\label{matrix}}
\end{figure}
To allow for the $A'$-type states, off-diagonal terms of competitive magnitude
must also be present.
These terms will act on configurations where a doubly occupied orbital and a singly
occupied orbital are separated by two empty sites (Fig. \ref{matrix}).
The off-diagonal matrix element leads to a new configuration 
where the doubly occupied site and the single particle
have moved past one another, essentially trading places
while conserving their center-of-mass.
We denote the strength of this matrix element by $-t$.
At the same time, since the particle configuration described here
involves three particles
in four adjacent sites, there must be an energy cost $V$
associated with it. It is easy to see that for
\be\label{Vt}
    V=t\;,
\ee
the $A'$-pattern is a zero energy eigenstate of the resulting Hamiltonian.
To see this, we first discuss the case of odd particle number.
Indeed, the matrix elements defined above immediately imply that there must
also be zero energy states for $N$ odd,\cite{read2}
 in the thin-torus limit. In this limit, these states are equal
amplitude superpositions of all states connected by the
off-diagonal matrix element described above, where 
a singly occupied orbital forms a defect between two adjacent
$A$-patterns, and becomes delocalized by means of the process  shown in Fig. \ref{matrix}. It is easy to see that for
a given spin state of the defect site, there
are two classes of states connected by such matrix elements.
This gives rise to two ground states for each value of $S_z$, related by a single magnetic translation.
The fact that the equal-amplitude superpositions of this kind are zero energy
eigenstates of the  Hamiltonian defined above can be understood
as a consequence of a ``detailed balance'' condition:
Consider a state entering the superposition, 
with the defect in a fixed
position.
The diagonal energy associated with this state ($2t$)
equals $t$ times the number of other such states ($2$) that have direct matrix elements
with the state under consideration. Note that the diagonal energy is indeed $2t$,
because of the doubly occupied orbitals to the left and right of the defect.
General solvable Hamiltonians  satisfying detailed balance 
conditions have been studied in Ref. \onlinecite{henley}.

The above considerations immediately extend to the $A'$-type states for
even particle number. Now each state in the equal-amplitude
superposition has two defects, which form a singlet. We see that the
detailed balance condition is still satisfied. The only states in the superposition
that
require additional consideration are those where the two defects are near
neighbors, as in the first line of Fig. \ref{Aprime}(b). 
In this configuration, the diagonal energy cost
is reduced by half, since each defect sees only one neighboring doubly
occupied site, three orbitals away. The other neighbor is given by the other defect,
two sites away. However, as stated above, two particles forming a singlet do not
repel each other, even at close distance. The energy of the state is thus
again $2t$. On the other hand, there are only two other configurations
(as opposed to four) with direct matrix elements into this state.

 We see now that the above matrix elements lead to the known
 thin-torus ground states. Interestingly, however, they also imply
 the existence of gapless excitations in the thin-torus limit.
 This can be seen in two different ways.
 First, we can consider an equal-amplitude superposition like the
 $A'$-states, but with the defect pair put into a triplet state.
 This will lead to a new orthogonal state with a finite energy expectation
 value. However, this energy will vanish in the thermodynamic limit,
 since the amplitude of each configuration scales as $1/N$, and
 the number of offensive configurations scales as $N$, leading
 to an energy expectation value of order $t/N$.
 
 Moreover, we can also consider giving the delocalized singlet
 defect in the $A'$-state a small but finite momentum $k$.
 It is easy to see that this leads to a dispersion proportional to $k^2$,
 in agreement with general arguments\cite{henley} for Hamiltonians
 with the detailed balance property.
 These findings seem consistent with the notion that the
 HR state can be thought of as a critical state between the strong- and 
 weak-pairing phases of a $d$-wave superconductor of composite fermions.\cite{readgreen}
 The implication of gapless excitations in the thin-torus limit of the hollow core Hamiltonian
 is the main result of this paper.

\section{Domain walls and zero-mode counting on the torus\label{count}}

\begin{figure}
\includegraphics[width=9cm]{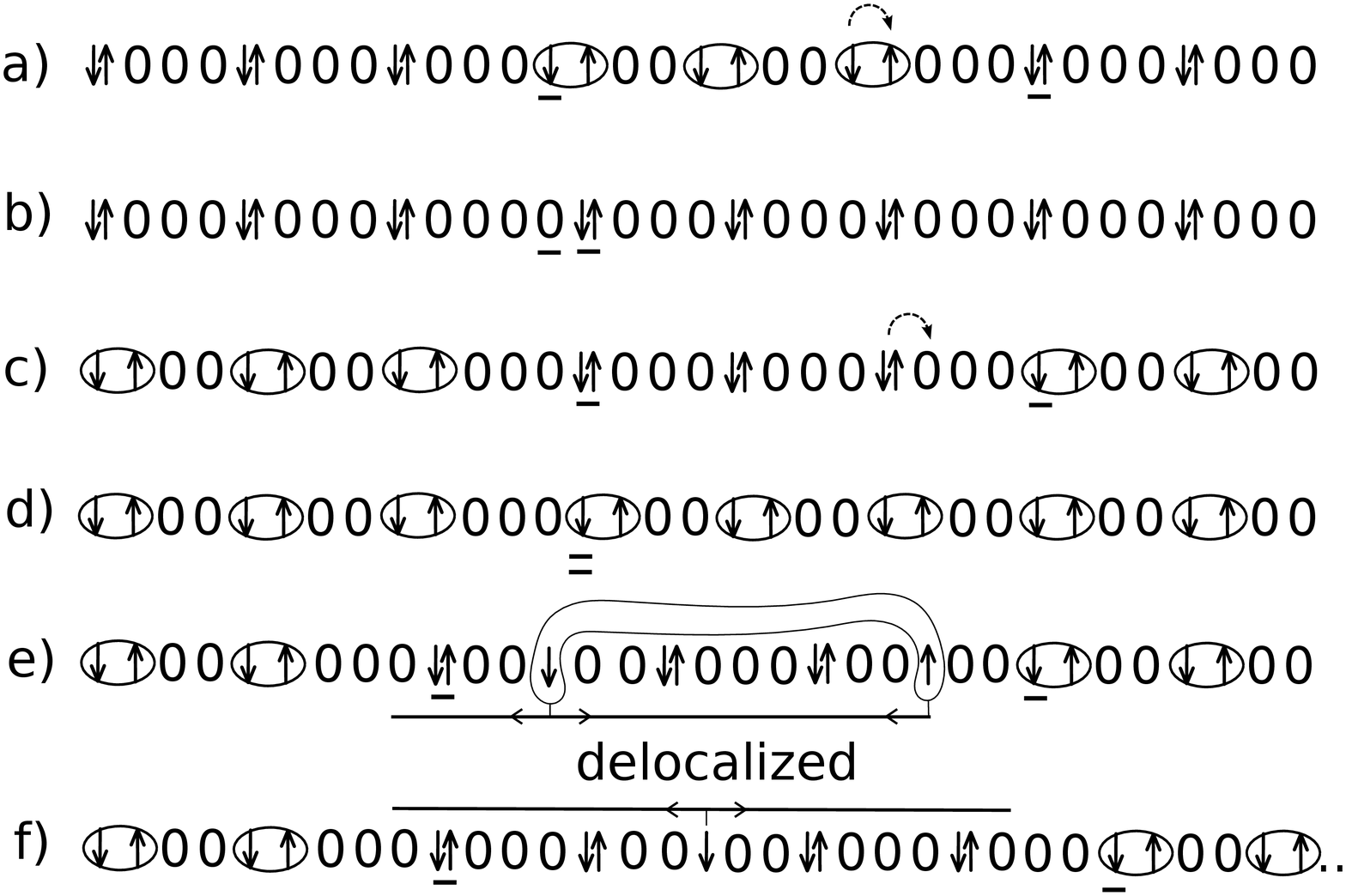}
\caption{
Charge-1/4 domain walls between $A$- or $A'$- and $B$-strings.
Underscores label the domain wall positions as defined in the text.
The hopping process indicated in (a) moves the right domain 
wall in the $ABA$ sequence to the left by four orbitals.
(b) shows the situation after three such moves, which 
shrink the size of the $B$-string to zero and lead to an 
additional $0$ between mutually shifted $A$-strings.
(c),(d) show the same for a $BAB$ sequence.
Note that for the conventions defined in the text,
the merged domain walls now coincide at the same orbital.
In any case, each allowed zero energy sequence of patterns can
be assigned a sequence of domain wall positions
satisfying \Eq{w} in a unique manner. (e) and (f)
show patterns involving $A'$-strings with 
two (e) or one (f) delocalized defect.
Only a ``snapshot'' of the state is shown with defects in
fixed positions, where it is understood that these
defects are delocalized as indicated and stated in the text.
The right defect in (e) is shown at the rightmost 
possible position.
\label{domain}}
\end{figure}

The zero modes of the hollow-core Hamiltonian
consist not only of the ground states at filling factor $\nu=1/2$,
but also of all states that have $2n$ quasi-hole type excitations added to a $\nu=1/2$
liquid. We are now in a position to identify the thin-torus limits of the complete
set of zero modes. Enumerating the number of zero mode states for fixed
 particle number $N$ and fixed $n$ has been an
integral part of the study of solvable Hamiltonians on the sphere,\cite{read2, ardonne1, ardonne2,ardonne3, read1}
where the counting is aided by the polynomial structure of the underlying wave functions.
Here we will generalize some of these results to the torus.
The thin-torus limiting states serve as natural bookkeeping devices for this task, enabling 
the counting of zero modes even in the absence of a simple polynomial structure of the
many-body wave functions. Note that, although we make use of the simplicity of the
thin-torus states for the counting, the number of zero modes does not depend on the aspect
ratio of the torus. Remarkably, the torus counting formulas we obtain in this way retain
the structure of their counterparts for the sphere, and do not involve additional sums
over topological sectors, as one might naively expect.

In the thin-torus limit, quasi-holes manifest themselves as domain walls between
the various ground-state patterns.\cite{seidel1, karlhede2}
In the $\nu=1/2$ state, the charge of elementary quasi holes is 1/4.\cite{HR}
Domain walls of this charge can be made between $A$- and $B$-type ground state
patterns as shown in Fig. \ref{domain}.
It is clear that these domain walls avoid any energy cost from the dominant
terms of the thin-torus Hamiltonian as described in the preceding section.
Any $A$-string pattern allows for the presence of a singlet pair of defects that
is delocalized within the string as in an $A'$-type ground state, Fig. \ref{domain}(e).
When this is the case we will continue to speak of an $A'$-string within the sequence of patterns.
It is easy to see that the detailed balance condition remains satisfied at the boundary
of such an $A'$- string. The reasons are similar to those discussed above for configurations
as shown in the first line of Fig. \ref{Aprime}(b),
 where the members of the singlet defect pair 
are close to each other.
 It follows that states containing $A'$-strings,
separated by hole-type domain walls from other permissible strings,
are zero energy eigenstates of the thin-torus Hamiltonian.
For the same reasons,
a single delocalized defect with given $S_z$ may be embedded into 
an  $A$-string, Fig. \ref{domain}f),
as in the odd $N$ ground states at $\nu=1/2$ discussed above.
We will also refer to such strings as $A'$-strings, with the 
understanding that $A'$-strings can harbor one or two delocalized defects.
To enumerate the number
of zero mode states for $N$ particles with $2n$ quasi-holes present,
we thus need to count the number of all string-sequences of the form
\be
   ABABA'BABA'\dotsc
\ee
with $2n$ domain walls, where strings can have variable lengths, including 0 (for the fusion of two
domain walls at the same point, see Fig. \ref{domain}(b),(d) ), and $B$-strings alternate with $A$- or $A'$-strings. In the following, we will always assume $n>0$, unless explicitly stated otherwise.

Knowing the structure of general quasi-hole states in the thin-torus limit,
we can count their number. We define the domain-wall position $w_i$, $0\leq i <2n$, as the position
of the first occupied orbital in an $A$-, $A'$-, or $B$-string\footnote{For an $A'$-string, this leading
position is that of the first doubly occupied site when the defect particle is not
at the left end of the string. It is greater by one than the left-most possible position
of the defect.}  (Fig. \ref{domain}). 
We first consider only a subset of patterns that satisfy a ``special boundary condition'' where $w_0=0$ is the leading position of an $A$- or $A'$-string.
Let $\Phi_0(N,n)$ be the number of such patterns for $N$ particles and $2n$ domain walls.
The LLL then consists of $qN+n$ orbitals, with $q=2$ in the present case. We will, however,
later generalize the result to $\nu=1/q$ HR states.
It is easy to see that,
if all patterns satisfying the special boundary condition are translated in all possible
$qN+n$ ways, each viable pattern is generated exactly $n$ times.
It follows that the total number of states for $N$ particles with $2n$ quasi-holes
is
\be\label{Phi}
\Phi(N,n,q)=\frac{qN+n}{n}\, \Phi_0(N,n)\;.
\ee
To find an expression for $\Phi_0(N,n)$, let $F$ be the total number of defects present
in the $A'$-strings of a pattern. Let $\Phi_{0;F}(N,n)$ be the number of patterns satisfying 
special boundary conditions and having $F$ defects. For each such pattern, we can consider
the pattern obtained by ``squeezing out'' all the defects. This is then a pattern with $N-F$ particles, and the same number of domain walls, all of which are between $A$- and $B$-strings, with no $A'$-strings present. The number of such patterns is $\Phi_{0;0}(N-F,n)$. Conversely,
to go from a pattern without defects and $N-F$ particles to one with $F$ defects and
$N$ particles, $F$ defects must be distributed over $n$ $A$-strings. Each $A$-string 
can either harbor one defect of any spin projection, or two defects forming a singlet.
Distributing the $F$ defects over $n$ $A$-strings is thus exactly the same problem
as putting $F$ spin-$1/2$ fermions into $n$ available spatial orbitals. The number of ways to
do this is $\left(
\begin{array}{c}
2n\\
F\\
\end{array}
\right)$, such that 
\be\label{Phi0F}
\Phi_{0;F}(N,n)=\left(
\begin{array}{c}
2n\\
F\\
\end{array}
\right)
\Phi_{0;0}(N-F,n)\;.
\ee
A similar combinatorial factor 
of $\left(
\begin{array}{c}
2(n-1)\\
F\\
\end{array}
\right)$
appears in the same problem on the sphere.\cite{read2}
The difference is owed to boundary conditions on the sphere,
which, in the absence of quasi-holes at the poles,
 enforce the presence of $A$-type strings 
at both boundaries in the ``dominance pattern'' of the state.
The latter might also be thought of as the ground state in
a ``prolate spheroid limit,'' where such boundary conditions 
should become manifest 
at a Hamiltonian level, 
in close analogy to the discussion given above
for the torus.\footnote{We are indebted to N. Read for this suggestion.}
 In particular, this boundary condition is the reason for the
non-degeneracy of the ground state at the ``incompressible''
filling factor on the sphere.

 It thus remains to calculate $\Phi_{0;0}(N,n)$.
 This is straightforward. 
 The possible domain wall positions are characterized by
 the conditions
 \be\label{w}
\begin{split}
 &0=w_0\leq w_1\leq\dotsc\leq w_{2n-1}<qN+n\\
 &w_{2j}=w_{2j-1}+1 \mod 2q\\
 &w_{2j+1}=w_{2j} \mod 2q\;.
 \end{split}
 \ee
 These can be simplified by introducing new integers $k_i$ via
 \be
 w_{2j}=2qk_{2j}+j\;,\;w_{2j+1}=2qk_{2j+1}+j\;.
 \ee
 The only constraint on the $k_i$ is then
 \be
 0=k_0\leq k_1\leq  k_2 \leq\dotsc \leq k_{2n-1}\leq \frac{N}{2}\,.
 \ee
  The number of possibilities to satisfy this constraint is just the number of (weak) compositions
  of $N/2$ into $2n$ parts, and is thus
  \be\label{Phi00}
   \Phi_{0;0}(N,n)=\left(
\begin{array}{c}
{N}/{2}+2n-1\\
2n-1\\
\end{array}
\right)\;.
  \ee
Putting together Eqs. \eqref{Phi0F}, \eqref{Phi00} , summing over all relevant values of $F$
and plugging into \Eq{Phi} gives
  \be\label{Phi_HR}
  \begin{split}
&\Phi_{\mbox{HR,torus}}(N,n,q)=\\
&\frac{qN+n}{n}\,\sum_{F,\, F=N\!\!\!\!\mod 2}\left(
\begin{array}{c}
2n\\
F\\
\end{array}
\right)
\,\left(
\begin{array}{c}
{(N-F)}/{2}+2n-1\\
2n-1\\
\end{array}
\right)\;.
\end{split}
\ee
This result 
is also valid for HR states at general filing factor $\nu=1/q$, $q\geq 2$.
 For this we only need to observe that the unit cells for the $A$- and $B$-patterns become
 \be
 \includegraphics[width=5.5cm]{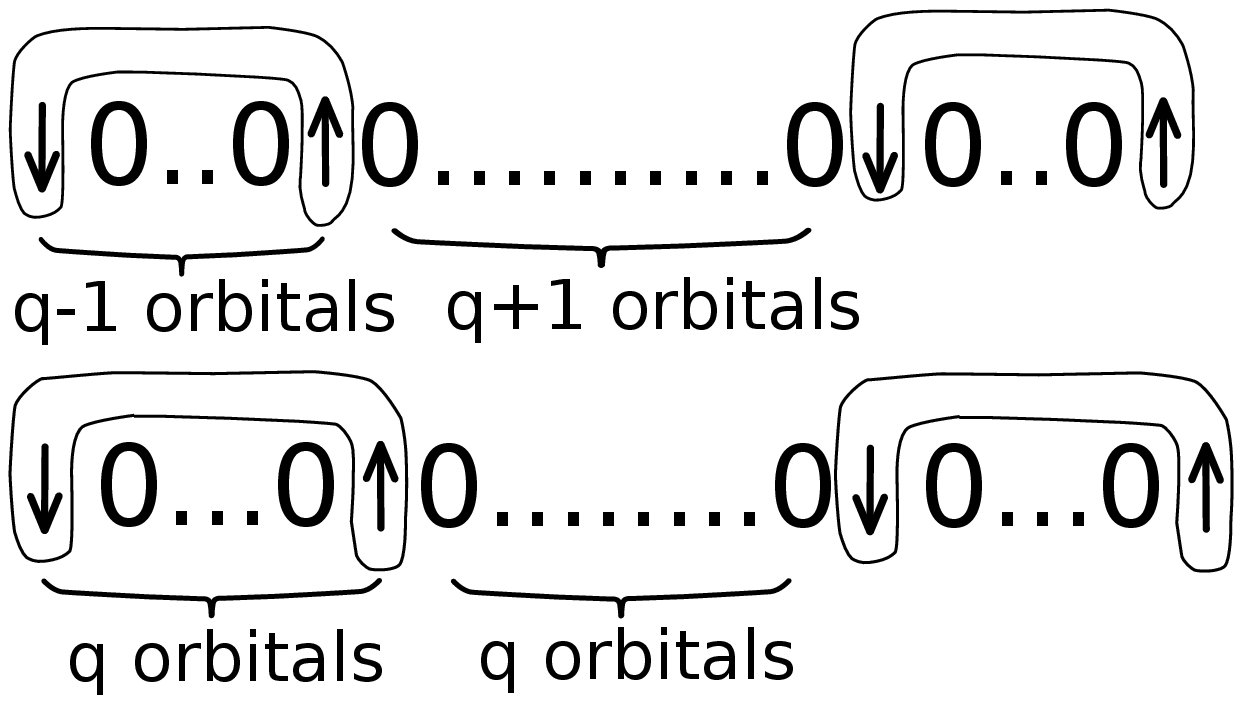}
 \ee
 at general $q$, which easily follows from a generalization of the calculation presented in the
 preceding sections, where $q$ replaces the exponent $2$ in the Laughlin-Jastrow factor.
  $A'$-strings likewise consist of delocalized charge neutral defects inside
 an $A$-pattern, subject to analogous rules. Only \Eq{Phi} depends on $q$, and the validity of its 
 generalization to general $q$ is obvious. 
 
 One may also be interested  in the number of the zero-mode multiplets of given total
 spin $S$, for given $n$, $N$, $q$. It is clear that this can be answered similarly, through a straightforward
 replacement of the combinatorial factor in \Eq{Phi0F} by that corresponding to the number
 of spin $S$-representations in a system of $F$ spin-$1/2$ fermions with access to
 $n$ states.
 
 With the ``special'' boundary condition replaced by the boundary conditions for the sphere that we
 identified above, the same method yields\footnote{The shift of $n$ by $1/2$ in the last term reflects the
 fact that we no longer fix the first of the domain wall positions to be at $0$.}
 \be\begin{split}\label{sphere}
 &\Phi_{\mbox{HR,sphere}}(N,n)=\\
& \sum_{F,\, F=N\!\!\!\!\mod 2}\left(
\begin{array}{c}
2(n-1)\\
F\\
\end{array}\right)
\Phi_{0;0}(N-F, n+1/2)\;,
 \end{split}
 \ee
in  agreement with Ref. \onlinecite{read2}.
This may serve as some confirmation that the matrix elements
of the thin-torus Hamiltonian have been identified correctly.  
Note that the torus result \eqref{Phi_HR} explicitly depends on $q$, whereas the 
one for the sphere does not.
  
  The strategy for zero mode counting presented here can be effortlessly
  generalized various other paired and related quantum Hall states.\footnote{For the Laughlin $\nu=1/q$ state on the torus,
  a simplified version of the argument yields 
 $\cfrac{Nq+n}{n} \left(
\begin{array}{c}
N+n-1\\
n-1\\
\end{array}
\right)
=\cfrac{Nq+n}{N} \left(
\begin{array}{c}
N+n-1\\
n\\
\end{array}
\right)
$
zero modes for $N$ particles with $qN+n$ flux quanta. The second expression holds for $n=0$ also.}
  For example, for the Moore-Read (MR) sequence of states, we can again
  distinguish
  two basic kinds of strings, which 
  may be separated by domain walls
  in the thin-torus limit of zero mode states.\cite{seidel2, karlhede3,ardonne4}
  One type of string must always contain an even number of particles, whereas the other
  may be even or odd in length. At $\nu=1/2$, e.g., we have $11001100\dotsc$,
  and $10101010\dotsc$,\cite{karlhede3}
  with the latter type of string being able to contain an odd  number of particles.
  This even/oddness can again be associated with a fermionic degree of freedom. \footnote{See Ref. \onlinecite{seidel_pfaff}
  for how this feature 
  of the Pfaffian patterns
  is related to the statistics of the state.} A calculation highly analogous to that carried out above for HR 
  states  yields
    \be\label{Phi_MR}
    \begin{split}
&\Phi_{\mbox{MR,torus}}(N,n,q)=\\
&\frac{qN+n}{n}\,\sum_{F,\, F=N\!\!\!\!\mod 2}\left(
\begin{array}{c}
n\\
F\\
\end{array}
\right)
\,\left(
\begin{array}{c}
{(N-F)}/{2}+2n-1\\
2n-1\\
\end{array}
\right)\;,
\end{split}
\ee
where $F$ corresponds to the number of spinless fermions associated with strings of odd length.\cite{seidel_pfaff}
The change in the fermionic combinatorial factor compared to \Eq{Phi_HR} reflects the fact that
the fermionic degrees of freedom are now spinless. The above may again be compared to the known result for the sphere.\cite{read2}  

Similarly, the MR sequence is intimately related\cite{greiter92_PRB} to the $(331)$-Halperin bilayer state.\cite{halperin1}
Here we adopt the convention\cite{read2} of referring to the Halperin state with $m=m'=q+1$, $n=q-1$
as a $(331)$-state at general filling factor $1/q$, the case $q=2$ being the proper $(331)$-state.
In the thin-torus limit,\cite{seidelyang} the $(331)$-patterns 
can be ``collapsed'' onto the MR patterns by dropping the (pseudo)-spin
information. At $\nu=1/2$, one pattern is $XX00XX00\dots$ where $XX$
denotes an equal-amplitude superposition between $\uparrow\downarrow$ and
$\downarrow\uparrow$. 
We will refer to this pattern as the ``$A$-pattern'', as it is also the pattern
associated with the ground state on the sphere.
This pattern collapses  onto the $11001100\dotsc$ MR pattern,
and combinatorially behaves in the same manner. The other MR-pattern is 
descended in the same way from what we will call the ``$B$-pattern'' of the $(331)$-state
\be
    \uparrow 0 \downarrow 0  \uparrow 0 \downarrow 0\dots   \uparrow 0 \downarrow \,.
\ee 
Here, we insist on the leading spin being $\uparrow$, and the final one being $\downarrow$.
More generally, we can of course add a sequence $\downarrow 0$ at the left end, and/or
a sequence $0\uparrow$ at the right end. These we can think of as the insertion of an up-spin
and/or down-spin fermion into a ``state'' provided by each $B$-string. This then exactly reproduces
the counting for the HR state, and we find that\footnote{We are indebted to E. Rezayi
for performing various checks on this relation.}
\be\label{Phi_331}
   \Phi_{\mbox{(331),torus}}(N,n,q)=   \Phi_{\mbox{HR,torus}}(N,n,q)\;.
\ee
We emphasize that this identity does not quite hold for the sphere,\cite{read2} chiefly due 
to the exchanged roles between $A$- and $B$-patterns with regard to harboring the 
spin-$1/2$ fermions.
On the torus, however, this exchange of roles is inconsequential.
We also note that \Eq{Phi_331} does not hold for $n=0$.
This is so because we can distinguish $B$-strings with $0$ and $2$
fermions only for $B$-strings that terminate in a domain wall. For $n=0$,
periodic boundary conditions on the torus prevent this distinction.

For the $(331)$-state we could also ask for the number of zero-modes with
given value of the total $S_z$. Again, this amounts to a trivial replacement  of the fermion
combinatorial factor in \Eq{Phi_HR}. 
The general relation \Eq{Phi_331} between the $(331)$ and HR counting 
also carries over to
sub-sectors of given total $S_z$.
Unlike for the HR case, however, 
the total spin $S$ is {\em a priori} not-well defined for the $(331)$-zero modes.

Finally, close connections between the HR state and the Haffnian\cite{wenwu,green}
have recently been observed.\cite{bernevig} There, a scheme has been proposed
to map the HR counting problem onto that of the Haffnian, by dressing the latter
 with spin according to certain rules.
In the present context, it is more natural to proceed along the reverse direction, which allows us
to obtain
an explicit Haffnian torus counting formula.
This can be done by applying the results obtained for the HR case to the Haffnian by ``dropping''
 the spin
degree of freedom.
More precisely,
we assume that
the matrix elements in the thin-torus limit of the Haffnian parent Hamiltonian\cite{green}
are the same as in the HR case, except that there is no analog
of the penalty for two particles  to be at distance $2$ or less, if they form
a triplet state. 
With this rule dropped, it is easy to see that now there is no limit
to the number of delocalized spinless defects that can be immersed into
the (now spin collapsed)  $A$-pattern. These defects should hence be
thought of as spinless bosons. Making the proper adjustments to
\Eq{sphere} reproduces the Haffnian counting on the sphere,\cite{green}
and making similar adjustments to \Eq{Phi_HR} yields
    \be\label{Phi_Haff}
  \begin{split}
&\Phi_{\mbox{Haff,torus}}(N,n,q)=\\
&\frac{qN+n}{n}\sum_{\substack{B\\
\, B=N\!\!\!\!\mod 2}}\left(
\begin{array}{c}
B+n-1\\
n-1\\
\end{array}
\right)
\,\left(
\begin{array}{c}
{(N-B)}/{2}+2n-1\\
2n-1\\
\end{array}
\right)\;.
\end{split}
\ee
This agrees with the data published in Ref. \onlinecite{bernevig}.
Again, this is valid for $n>0$. For $n=0$,
the non-trivial Haffnian torus degeneracy can be treated separately,\footnote{The number
of bosons that can enter an $A$-string is limited only by the length of that string.
This leads to an extensive ground state degeneracy\cite{rezayi_unpublished} even at $\nu=1/2$.}
 and
 has been discussed
from the point of view of dominance patterns recently.\cite{bernevig}

\section{Discussion\label{dis}}
  
In this work, we have analyzed the thin-torus limit of the Haldane-Rezayi state,
by first analyzing the behavior of the ground state wave functions, and 
inferring the dominant matrix elements of the thin-torus hollow-core Hamiltonian.
The latter were constructed to be consistent 
with the uniqueness of the ten zero energy ground states at filling factor $1/2$,
and with the necessary locality of the thin-torus Hamiltonian.
From there we were able to deduce the existence of both singlet and triplet
gapless excitations in the thin-torus limit.
Assuming adiabatic continuity of the low energy sector between the thin-torus
and  the 2D limit, the existence of gapless excitations in the
HR state follows, in agreement with general arguments.\cite{read_chiral, read_viscosity,read_preprint2}
So far the notion of adiabatic continuity has been mainly
studied in cases where an energy gap is believed to exist in the 
2D regime, in agreement with the thin-torus limit.\cite{seidel1,seidel2,karlhede2,karlhede3}
The findings made here do, however, have much in common
with our earlier analysis of a critical point between the (331)-state
and the MR state, using the thin-torus approach.\cite{seidelyang}
There we found gapless excitations in the thin-torus limit at a 
critical point, within one topological sector of the thin-torus theory,
  but not in others. We note that even in the fully gapped case,
  the excitation spectrum within different topological sectors is generally not identical in the thin
  torus limit.
  This is so because these sectors become locally distinguishable in this limit,
  and need not be related by any symmetry. Indeed, the ground state degeneracy itself
  could be lifted by a local perturbation away from the special solvable point.
 These spectral differences all disappear when we leave the thin-torus limit
 and cross over into the 2D regime. For these reasons we argued in Ref. \onlinecite{seidelyang}
 that gapless states may generically reveal their gapless excitations in the 
 $L_y\rightarrow 0$ limit of some topological sectors, but not of others.
 The present analysis strengthens this case, in particular because
  the same observations could be made for a solvable Hamiltonian,
 whereas in Ref. \onlinecite{seidelyang} we had to move away from the solvable point.
 
A small point worth mentioning is the fact that even though the $B$-type states
do not seem to couple locally to gapless excitations when the $L_y\rightarrow 0$
limit is taken, they do so in the ``dual'' limit $L_x\rightarrow 0$ of the torus.
In particular, a certain linear combination of $B$-ground states 
(i.e., states approaching $B$-patterns in the $L_y\rightarrow 0$ limit)
will approach $A$-type patterns in this dual limit (cf. Ref. \onlinecite{seidel2010}).
It is thus true that each of the states in \Eq{AB}
locally couples to
gapless excitations in at least one of the two mutually dual limits.
This somewhat removes the inequivalence
between sectors that exists when the $L_y\rightarrow 0$ limit is considered by itself.

Despite these subtleties, we argue that the presence of gapless excitations in 
the thin-torus limit can actually be a stronger indication for the nature of the 2D limit
than the absence of such excitations. To make this case, we consider a cylinder
with $L_x=\infty$ and finite $L_y$. If there are gapless excitations in the $L_y\rightarrow 0$
limit, a sufficient assumption is that for any finite $L_y$, the properties of the system
are analytic in $L_y$. This seems reasonable for the ``special'' solvable Hamiltonians.
However, if there are gapless excitations at any value of $L_y$, they exist by definition
in the limit $L_y\rightarrow \infty$, so long as this limit is well-defined. The converse, however,
is not true. Having an energy gap for any finite value of $L_y$ does not necessarily
prevent this gap from closing as the limit $L_y\rightarrow \infty$ is taken.\footnote{This is precisely
what we expect to happen in the $B$-sectors.}
The assumption of analyticity in $L_y$, while still subject to detailed
justification, is thus more powerful when gapless excitations are identified
in the thin-torus/cylinder limit.

{
We close this section with some comments on possible generalizations to other states, such as the gaffnian.\cite{gaffnian}
In general, we expect that the existence of gapless excitations cannot
be inferred from knowledge of the thin-torus patterns alone. 
Instead it requires the detailed study of a given parent Hamiltonian. 
It is likely a special feature of the HR state that the complete set of thin-torus patterns already contains enough information 
about the parent Hamiltonian such that the existence
of gapless excitations can be concluded (in the thin-torus limit). 
While in general a more direct analysis of the parent Hamiltonian is necessary, we are encouraged to believe that such an analysis is not only 
technically possible in the thin-torus limit but will yield results that are qualitatively correct also away from this limit. }


{
In this context, it may be of interest that various families of states have been discussed in the literature\cite{estienne,simon2}
that were proposed to have the same patterns of zeros
(as defined in Ref. \onlinecite{wenwang}). 
In particular, the family of $S_3$ states\cite{simon2} 
includes both unitary and non-unitary wave functions,
and parent Hamiltonians for both kinds have been proposed.
Two states in this infinite family can, in general, be distinguished by
their torus degeneracy, which presumably can become arbitrary
large. At the same time, at the incompressible filling factor $\nu=3/4$
there exist only 20 thin-torus patterns
that satisfy the generalized Pauli principle (GPP) associated
with the patterns of zeros (no more than three particles in
any four adacent orbitals). For an $S_3$ state whose torus degeneracy
$D\leq 20$, we expect the thin-torus limits of the ground states to be some subset 
of these 20 patterns. On the other hand, for $D>20$ it is necessarily possible
to form linear combinations of ground states that become orthogonal to these
20 patterns in the thin-torus limit. The thin-torus limits of such states
will be more complicated, and will violate the GPP. This may happen, e.g.,
in much the same manner in which our $A'$ states violate the GPP enforced by the
diagonal matrix elements in the thin-torus limit.
For these reasons, we find it not unlikely that thin-torus limits, taken for a complete orthogonal
set of torus ground states, may still distinguish different members of the $S_3$ family of states.
In particular, we see no obvious reasons why the adiabatic continuity conjecture must 
fail in those cases where a suitable parent Hamiltonian does exist.
Though complicated thin-torus limits may appear, 
 the case of the HR state presented here
suggests that such limits still 
contain valuable information. 
We believe that
a detailed investigation of the 
utility of the present approach to the $S_3$ states 
would be interesting, and leave this possibility for 
future study.
}

\section{Conclusion}

We have constructed the thin-torus picture of the Haldane-Rezayi state,
and found the existence of gapless excitations in this limit.
This provides some further evidence in favor of the conjectured gapless
nature of this state.
We have also used our results to derive, for the first time, zero mode
counting formulas for the special Hamiltonian of the HR state --and several others--
on the torus.
The observations made here in the thin-torus limit may serve to guide
a more rigorous proof of the gapless nature of the HR state.
For this one would need to identify
wave function expressions that analytically continue the gapless
excitations found here, say, to a torus of arbitrary size and aspect ratio.
We leave this exciting possibility for future work.

\begin{acknowledgments}
We are indebted to E. Ardonne, N. Read, E. Rezayi, and  S. Simon for insightful discussions. 
AS would like to thank R. Thomale for helpful comments on the manuscript.
This work was supported  by the National Science Foundation under NSF Grant No. DMR-0907793 (AS)
and NSF Grant No. DMR-1004545 (KY).
\end{acknowledgments}

\appendix

\section{The structure of the $p$-matrix\label{app1}}

In this and the following appendices, we present the remaining details 
in the calculation of the thin-cylinder limits of the HR ground states.
We first proove that the matrix \pab that leads to the maximum of \Eq{S} is always of the form
\eqref{pab}. To this end, we plug \Eq{na} into \Eq{S}, obtaining
\be\label{Sapp1}
S=\mbox{const} + \sum_{\alpha,\beta,\gamma} (p_{\alpha\beta}p_{\alpha\gamma}
+p_{\alpha\beta}m_{\alpha\gamma}+p_{\alpha\gamma}m_{\alpha\beta})\,.
\ee
For fixed indices $\alpha$, $\beta$, we extract the dependence of $S$ on the matrix element $p_{\alpha\beta}=-p_{\beta\alpha}$ via
\be\label{Spab}
  S=2(p_{\alpha\beta}^2+p_{\alpha\beta}r_{\alpha\beta})+\mbox{const}\;,
\ee
where the constant and $r_{\alpha\beta}$ depend on other matrix elements of the $p$-matrix, but not on \pab, and
\be\label{rab}
r_{\alpha\beta}=\sum_{\gamma,\gamma\neq\beta} (p_{\alpha\gamma} + m_{\alpha\gamma})
-\sum_{\gamma,\gamma\neq\alpha}(p_{\beta\gamma} + m_{\beta\gamma})\;.
\ee
It is clear from the quadratic structure of \Eq{Spab} that for any $p$-matrix that maximizes $S$,
$|\pab|$ must take on the maximum possible value $m_{\alpha\beta}$.
For otherwise, the value of $S$ could certainly be increased.
More precisely, it follows that 
\be
p_{\alpha\beta}=m_{\alpha\beta}\, \mbox{sign}(r_{\alpha\beta})\;,
\ee
unless it happens that $r_{\alpha\beta}=0$, in which case either sign of \pab is possible.
In either case, $\tilde r_{\alpha\beta}=r_{\alpha\beta}+2\pab$ is certainly non-zero
so long as $|p_{\alpha\beta}|=m_{\alpha\beta}$ is non-zero, and we always have
\be
p_{\alpha\beta}=m_{\alpha\beta} \,\mbox{sign}(\tilde r_{\alpha\beta})\,.
\ee
From \Eq{rab}, we also have
\be
\tilde r_{\alpha\beta}=r_\alpha-r_\beta\;,
\ee
where
\be
r_\alpha = \sum_\gamma (p_{\alpha\gamma}+m_{\alpha\gamma})\;.
\ee
There is a permutation $\rho$ of $N$ objects such that 
$r_{\alpha}\leq r_{\beta}$ for $\rho_\alpha<\rho_\beta$,
where equality  of $r_\alpha$ and $r_\beta$ can, by construction,
only hold for $m_{\alpha\beta}=0$.
For such a permutation, we then have
\be\label{pab2}
\pab= m_{\alpha\beta}\,\sign(\rho_\alpha-\rho_\beta)\;,
\ee
which is \Eq{pab}.
We have thus shown that this equation must hold for any $p$-matrix that leads to a
maximum of $S$, for some permutation $\rho$. As explained in the main text,
$\rho$ can be thought of as giving rise to an arrangement of particles
on a ``squeezed lattice''.

\section{The squeezed lattice: Effective Coulomb interaction\label{app2}}
The proper choice of $\rho$ depends on the symmetric matrix $m_{\alpha\beta}$,
which depends both on the state and on the ``pairing permutation'' $\sigma$ under consideration, as
defined in the main text, Eqs. \eqref{sconstraint}, \eqref{sconstraint2}.
Here we show that the proper choice of $\rho$, i.e. the arrangement of the particles
on the squeezed lattice, is obtained by minimizing an energy due to an effective 
attractive ``Coulomb interaction'' between members of a pair. To see this,
we first observe that in the second term of \Eq{Sapp1}, the sum over $\gamma$ gives rise to the $\alpha$-independent constant
$\sum_\gamma m_{\alpha\gamma}$, such that the term vanishes after summing over $\alpha$ and $\beta$
by symmetry. For the same reasons, the third term vanishes as well. We did not use this fact in
Appendix \ref{app1}, since in its present form, the argument given there can be used in more general situations
where $\sum_\gamma m_{\alpha\beta}$ depends on $\alpha$. This is the case, e.g., in Ref. \onlinecite{seidelyang}.
For the first term of \Eq{Sapp1}, we plug in a general relation of the form \eqref{mab} :
\be\label{Smab}
S=\sum_{\alpha\beta\gamma} s_{\alpha\beta}s_{\alpha\gamma}\,\mbox{sign}(\rho_\alpha-\rho_{\beta})
\mbox{sign}(\rho_\alpha-\rho_{\gamma})
+\mbox{const}\,.
\ee
According to \Eq{pab2}, for the $A$- and $B$-states, the $s$-matrix is just the $m$-matrix 
defined in Eqs. \eqref{sconstraint} and \eqref{sconstraint2}.
Here we will consider a slightly more general problem, of which these equations are
special cases. Consider a pairing of $N$ indices, where we denote the partner of the index $\alpha$
by $\overline{\alpha}$, such that $\overline{\overline{\alpha}}=\alpha$. Consider then a symmetric matrix
$s_{\alpha\beta}$ defined as
\be\label{sconstraint3}
\begin{split}
s_{\alpha\beta}=0 \quad&\mbox{for } \alpha=\beta\\ 
 s_{\alpha\beta}\equiv s_\alpha=s_{\overline\alpha} \quad &\mbox{for } \beta=\overline{\alpha}\\
  s_{\alpha\beta}=s\quad&\mbox{otherwise.}
\end{split}
\ee
Plugging this into \Eq{sconstraint3} yields, after some amount of straightforward calculation:
\be\label{coulomb}
-S=2\sum_\alpha s(s-s_\alpha) |\rho_\alpha-\rho_{\overline\alpha} |+\mbox{const}.
\ee
In the $A$- and $B$-state, $s_\alpha$ is a constant, and the coefficient
$s(s-s_\alpha)$ is positive (cf. Eqs. \eqref{sconstraint}, \eqref{sconstraint2}).
Note also that $\rho_\alpha$ and $\rho_{\overline\alpha}$ are the positions of
a pair on the squeezed lattice. We can thus interpret 
\Eq{coulomb} as the total energy of particles on the squeezed lattice
where pairs interact via an attractive linear 1D Coulomb potential.
Clearly, minimization of this energy requires one to place members of a pair
onto adjacent sites of the squeezed lattice, as shown in \Eq{rho}.
For the $A$- and $B$-states, this then has the implications stated in the main text.

The above observations are now easily extended to the $A'$-states.
The structure of the $s$-matrix was worked out in the main text, \Eq{mab2}.
Recall that a single pair has been excluded from the pairing $\cal P$.
\Eq{mab2} is then another special case of \Eq{sconstraint3},
with $s_\alpha=0$ for all except one pair of indices, for which
$s_\alpha=s_{\overline\alpha}=1=s$. The implications of this are plainly apparent in
\Eq{coulomb}. All pairs attract in the same manner as before, except for the
special ``broken'' pair, whose Coulomb attraction has now been switched off.
The position of the members of the broken pair on the squeezed lattice are then
arbitrary; they all lead to the same maximum value of $S$. All the other
pairs must still be nearest neighbors on the squeezed lattice.
The resulting ``un-squeezed'' thin-cylinder state is the equal-amplitude superposition
described in the bulk of the paper.

The present method to work out the thin-torus patterns of the Haldane-Rezayi
state can be used for other paired states as well. In particular, it applies
almost without change to Moore-Read states. This, in addition to the
similar structures of the counting rules found here and in Ref. \onlinecite{read2},
is another manifestation of various formal connections between these states.

\section{Coefficient of the dominant terms in the $A'$-state polynomial\label{app3}}

Here we calculate the coefficient of the dominant monomials 
in the $A'$-state thin-cylinder limit, as identified in Appendix \ref{app2}. 
These monomials correspond 
to a $p$-matrix
of the form defined in Eqs. \eqref{mab}, \eqref{mab2}. In particular, this will prove that these coefficients
are non-zero. Recall that $\cal P$ in \Eq{mab2} is a pairing of the particle indices into $N/2-1$
pairs, with one pair left out. The permutation $\rho$ must give rise to a squeezed lattice
configuration with all pairs in $\cal P$ nearest neighbors. This is necessary to maximize
$S$, and hence the $\kappa$-dependent part of the amplitude \eqref{amplitude}, as shown in 
Appendix \ref{app2}. All monomials corresponding to a $p$-matrix satisfying these rules can be obtained
only from this particular $p$-matrix. This is so because other $p$-matrices satisfying these
rules will either permute particle indices with like spins (which for fermions, must always lead to a different monomial).
Or, in general, other such $p$-matrices lead to different states with the ``defect'' particles
occurring in the $A'$-pattern in different positions. On the other hand, $p$-matrices not 
satisfying the rules above will lead either to monomials with 
vanishing coefficients or to ones with lower $S$.
For these reasons, we only need to focus on a particular $p$-matrix
satisfying the rules summarized here, and calculate the coefficient of the monomial
obtained by choosing the term $\xi_\alpha^{m_{\alpha\beta+\pab}}
\xi_\beta^{m_{\alpha\beta-\pab}}$ in factors depending on
$\xi_\alpha$ and $\xi_\beta$ in \Eq{HRC}.
It is easy to see that all such $p$-matrices lead to a coefficient of the same value, up to a sign.
We thus choose the pairing of up-spin and down-spin indices given by
\be
{\cal P}= \{(2\downarrow, 2\uparrow) , (3\downarrow, 3\uparrow) \dotsc (N'\downarrow, N'\uparrow)\}\,,
\ee
where we write $N'=N/2$ for convenience, and
 the pair $(1\uparrow, 1\downarrow)$ is left out. This  defines an $s$-matrix according to
\Eq{mab2}. We also choose a squeezed lattice configuration, $\rho$, consistent with this 
pairing (Appendix \ref{app2}), and define \pab from \Eq{mab}.

Now we first fix the permutations $\sigma$, $\lambda$ in \Eq{HRC} and generate a monomial
using the $p$-matrix from the resulting polynomial of the form \eqref{pseudoLJ}, where the signs
depend on $\sigma$ and $\lambda$.
The coefficient of the resulting monomial is of the form
\be
(-2)^{N'-1} \chi(\sigma,\lambda)\;,
\ee
where the first term indicates that $N'-1$ of the mixed terms $\xi_\alpha\xi_\beta$ are chosen, 
one for each pair in $\cal P$, except that not all of them have a coefficient $-2$. Instead, some
contribute $+2$, and this change of sign is accounted for in the factor
\be\label{chi}
  \chi(\sigma,\lambda)= (-1)^{\sigma+\lambda}(-1)^{g(\sigma,\lambda)}\;,
 \ee
 where
 \be
 g(\sigma,\lambda)=\sum_{r=2}^{N'} {\delta_{\sigma_r,\lambda_r}(1-\delta_{\sigma_r,1})}\,.
 \ee
 is just the number of pairs $(\sigma_r\downarrow, \lambda_r\uparrow)$ with $r>1$ that are
 contained in
 $\cal P$.
The full coefficient of the resulting monomial then satisfies:
\be\label{coeff2}
C_{\{n_\alpha\}}=(-2)^{N'-1} \sum_{\sigma ,\lambda \in S_{N'}} \chi(\sigma,\lambda)\,.
\ee
To evaluate this, we represent $\chi(\sigma,\lambda)$ through a binomial sum:
\be
  \chi(\sigma,\lambda)=\sum_{n=0}^{N'-1} (-2)^n\, \chi_n(\sigma,\lambda)\,,
\ee
where
\be\label{chin}
  \chi_n(\sigma,\lambda)=(-1)^{\sigma+\lambda}\left(
\begin{array}{c}
g(\sigma,\lambda)\\
n\\
\end{array}
\right)\,.
\ee
This is obtained from \Eq{chi} by simply applying the binomial theorem to $(-1)^g=(1-2)^g$,
and observing that $g(\sigma,\lambda)\leq N'-1$.
The point of doing this is that one may see that
\be\label{chin0}
\sum_{\sigma ,\lambda \in S_{N'}} \chi_n(\sigma,\lambda)=0\mbox{ for } n<N'-1\;.
\ee
To see this, we cast \Eq{chin} in the form
\be\label{chin2}
  \chi_n(\sigma,\lambda)=(-1)^{\sigma+\lambda}\sum_{\substack{~\omega \in 2^{\Gamma}\\
   |\omega| =n}} \,\prod_{r\in\omega} \delta_{\sigma_r,\lambda_r}
  \,,
\ee
where $\Gamma=\{2,3,\dots,N'\}$, $2^\Gamma$ is the set of all subsets of $\Gamma$, 
and the sum is over all subsets $\omega$ with $n$ elements.
Clearly for given $\sigma$, $\lambda$ there are $\left(
\begin{array}{c}
g(\sigma,\lambda)\\
n\\
\end{array}
\right)$ distinct choices for $\omega$ that contribute to the sum.
However, for $\omega$ {\em fixed} and $n<N'-1$ we may consider
defining a new permutation $\tilde\sigma$ obtained from $\sigma$
by exchanging the values of $\sigma_{r_1}$, $\sigma_{r_2}$ for the
smallest two indices $r_1$, $r_2$ that are {\em not} contained in $\omega$.
It is clear that when summing \Eq{chin} over $\sigma$, terms with
given $\omega$ and $\sigma$, $\tilde\sigma$ so related will cancel.
Thus \Eq{chin0} follows.
Eqs. \eqref{chi} and \eqref{chin0} in \Eq{coeff2} yield
\be
C_{\{n_\alpha\}}=(-2)^{2N'-2} \sum_{\sigma ,\lambda \in S_{N'}} \chi_{N'-1}(\sigma,\lambda)\,.
\ee
By definition of $\chi_{N'-1}(\sigma,\lambda)$, however, the pairs $\sigma$, $\lambda$ contributing
to this last sum are exactly those with $\sigma=\lambda$ and $\sigma_1=\lambda_1=1$.
There are $(N'-1)!$ such pairs. This gives
\be\label{coeff_result}
C_{\{n_\alpha\}}=2^{N-2} (N/2-1)!
\ee
as stated in the main text, which is non-zero.

\bibliography{hr}

\end{document}